\documentclass[twocolumn,showpacs,preprintnumbers,amsmath,amssymb]{revtex4}

\usepackage{graphicx}
\usepackage{bm}
\usepackage{amsmath}

\begin{document}

\title{Giant Anisotropy of Spin-Orbit Splitting at the Bismuth Surface}

\author{Y. Ohtsubo$^1$, J. Mauchain$^2$, J. Faure$^3$,
E. Papalazarou$^2$, M. Marsi$^2$, P. Le F\`evre$^1$, F. Bertran$^1$, A. Taleb-Ibrahimi$^1$, L. Perfetti$^{4}$}

\affiliation{%
$^{1}$ Synchrotron SOLEIL, L'Orme des Merisiers, Saint-Aubin-BP 48, F-91192 Gif sur Yvette, France}
\affiliation{$^{2}$ Laboratoire de Physique des Solides, CNRS-UMR 8502, Universit\'e Paris-Sud, FR-91405 Orsay, France}
\affiliation{
$^{3}$ Laboratoire d'Optique Appliqu\'ee, Ecole polytechnique, 91128 Palaiseau cedex, France}
\affiliation{$^{4}$ Laboratoire des Solides Irradi\'{e}s, Ecole polytechnique, 91128 Palaiseau cedex, France}

\begin{abstract}

We investigate the bismuth (111) surface by means of time and angle resolved photoelectron spectroscopy. The parallel detection of the surface states below and above the Fermi level reveals a giant anisotropy of the Spin-Orbit (SO) spitting. These strong deviations from the Rashba-like coupling cannot be treated in $\textbf{k}\cdot \textbf{p}$ perturbation theory. Instead, first principle calculations could accurately reproduce the experimental dispersion of the electronic states. Our analysis shows that the giant anisotropy of the SO splitting is due to a large out-of plane buckling of the spin and orbital texture.

\end{abstract}

\maketitle

The realization of transistors via the transport of spin polarized electrons has attracted the interest of the solid state community since 20 years \cite{Datta,Zutic}. In such devices, an applied gate voltage induces a spin torque of the injected electrons via Spin-Orbit (SO) interaction. The energy scale of this effect is typically 1-10 meV in the semiconductor heterostructures \cite{Nitta}, but reaches values 10 times larger at surfaces of systems containing heavy elements \cite{Lashell,Koroteev,Koroteev1,Ast}. Therefore, the latter are considered as valuable models for future spintronic applications \cite{Crepaldi}. This cross fertilizing field has been recently enriched by the discovery of protected edge states in topological insulators \cite{Hasan}.

The effects of SO coupling at the surface of solid states materials have been reviewed by several authors \cite{Hofmann,Hugo,Hasan1}. As originally noticed by Rashba, the spin degeneracy of the electronic states is lifted by the breakdown of inversion symmetry \cite{Rashba}. Being a relativistic effect, the SO splitting  arises from the asymmetry of the electronic wavefunction in proximity of the ionic cores \cite{Bihlmayer}. Despite the complexity of this problem, the $\textbf{k}\cdot \textbf{p}$ perturbation theory provides the leading terms of the SO Hamiltonian for small electronic wavevectors. When the surface has $C_{3v}$ symmetry, the first order term is indeed the Rashba hamiltonian $\alpha_R(k_y\sigma_x-k_x\sigma_y)$. This interaction term leads to an isotropic spin splitting and chiral spin texture. It reproduces correctly the Shockley states at the (111) surface of gold as well as the electronic properties of several heterostructures \cite{Lashell,Rashba,Hoesch}.
Nonetheless, higher order expansions become necessary in systems where the surface state cannot be modeled within the framework of the nearly-free-electron approximation.
As an example, the topological insulators Bi$_2$Te$_3$ and Bi$_2$Se$_3$ show a trigonal warping which is linked to the third order term $k_x(k_x^2-3k_y^2)\sigma_z$ \cite{Fu,Souma,Kuroda}. As a consequence, the spin polarization acquires an out-of-plane component \cite{Claessen} that would affect the spin transport in non-ballistic devices \cite{Cartoixa}.

In this letter we show that the large deviations from the Rashba hamiltonian induce a giant anisotropy of the SO splitting at the (111) Bismuth surface. We make use of an ultrashort laser pulse to efficiently populate electronic states up to 0.5 eV above the Fermi Level. By these means, we observe that the SO splitting is $\Delta_{\Gamma M}=150\pm10$ meV along the $\Gamma$-M direction and increases by 250\% for a rotation of $6^\circ$ around the surface normal. We checked the SO anisotropy by ab initio calculations of the band structure, finding excellent agreement with the observed dispersion of electronic states. The analysis of the resulting wavefunctions indicates that the SO anisotropy is due to the buckling of the spin orientation out of the surface plane. By projecting the spin-polarized wave functions on a local basis, we find an out-of-plane component larger than 30\%. Our result differs from previous DFT calculations \cite{Zhang} whereas corroborates the spin resolved measurements of the occupied electronic states \cite{Takayama}. In order to test the entanglement between the spin and orbital degrees of freedom  \cite{Park}, we also calculated the projections of the Kohn-Sham wavefunctions on the atomic orbital set. It follows that spin and orbitals have an opposite and nearly proportional polarization.

Photoelectron spectra with photon energy of 18 eV have been collected at the Cassiop\'ee beamline of Soleil Synchrotron. The sample has been measured at 20 K with energy resolution of 10 meV and angular resolution of 0.2 degrees. The bismuth (111) surface has been prepared by sputtering-annealing cycles of a polished monocrystal.
Time resolved photoelectron spectroscopy experiments have been performed with the FemtoARPES setup, using a Ti:Sapphire laser that generates 35 fs pulses centered at 790 nm with repetition rate of 250 kHz.
Part of the beam is employed to generate the fourth harmonic by a cascade of frequency mixing in BBO crystals ($\beta$-BaB$_2$O$_4$) \cite{Faure}. The 197.5 nm probe and the 790 nm pump are focused on the sample with a spot diameter of 100 $\mu$m and 200 $\mu$m, respectively. Their cross-correlation in a BBO crystal has a full width at half maximum (FWHM) of 80 fs. The bandwidth of the 197.5 nm beam (6.3 eV) limits the overall energy resolution of TRPES spectra to 60 meV. All the time resolved measurements have been performed with incident pumping fluence of 0.6 mJ/cm$^2$ and sample temperature of 130 K.

The DFT calculations have been done using the \mbox{\lq\lq APW + lo \rq\rq} method of the WIEN2k code \cite{Blaha} with the SOI taken into account and the ex-corr. functional based on GGA \cite{Perdew}. The surface was modelled by a symmetric slab of 20 Bi layers repeated along [111] with a periodic gap of 10 \AA~~ \cite{Koroteev,Koroteev1}.
The interlayer distances in the slab were set to the values determined by the X-ray diffraction
tecnique \cite{Cucka62}.

\begin{figure} \begin{center}
\includegraphics[width=1\columnwidth]{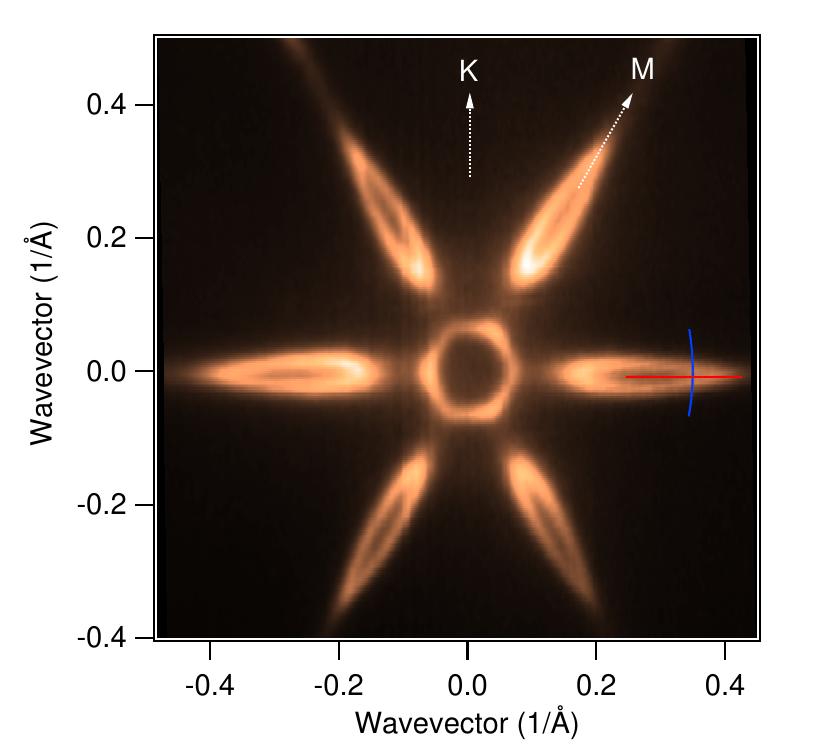}
\caption{Photoelectron intensity map integrated in an energy window of 20 meV around the Fermi level. The data have been acquired with circularly polarized photons centered at 18 eV. The red and blue line stand for the wavevectors cut of Fig. \ref{Fig3}C,D and Fig. \ref{Fig4}A,B, respectively.} \label{Fig1}
\end{center}
\end{figure}

Figure \ref{Fig1} shows a photoelectron intensity map acquired with 18 eV photons in an energy window of 20 meV centered at the Fermi level. The crossing points of surface states generate an internal pocket of hexagonal shape and 6 elongated lobes along the $\Gamma$-M direction. The resulting map is in excellent agreement with the original measurements of Ast \textit{et al.} \cite{Ast0} and bears no resemblance to the Fermi surface of nearly-free-electron systems \cite{Lashell}.
In a following work, Korotheev \textit{et al.} have reproduced the experimental band structure by performing first principle calculations that accounted for the SO coupling \cite{Koroteev,Koroteev1}. By these means, they proved that SO splitting has a major effect on the electronic properties \cite{Note}.

\begin{figure} \begin{center}
\includegraphics[width=1\columnwidth]{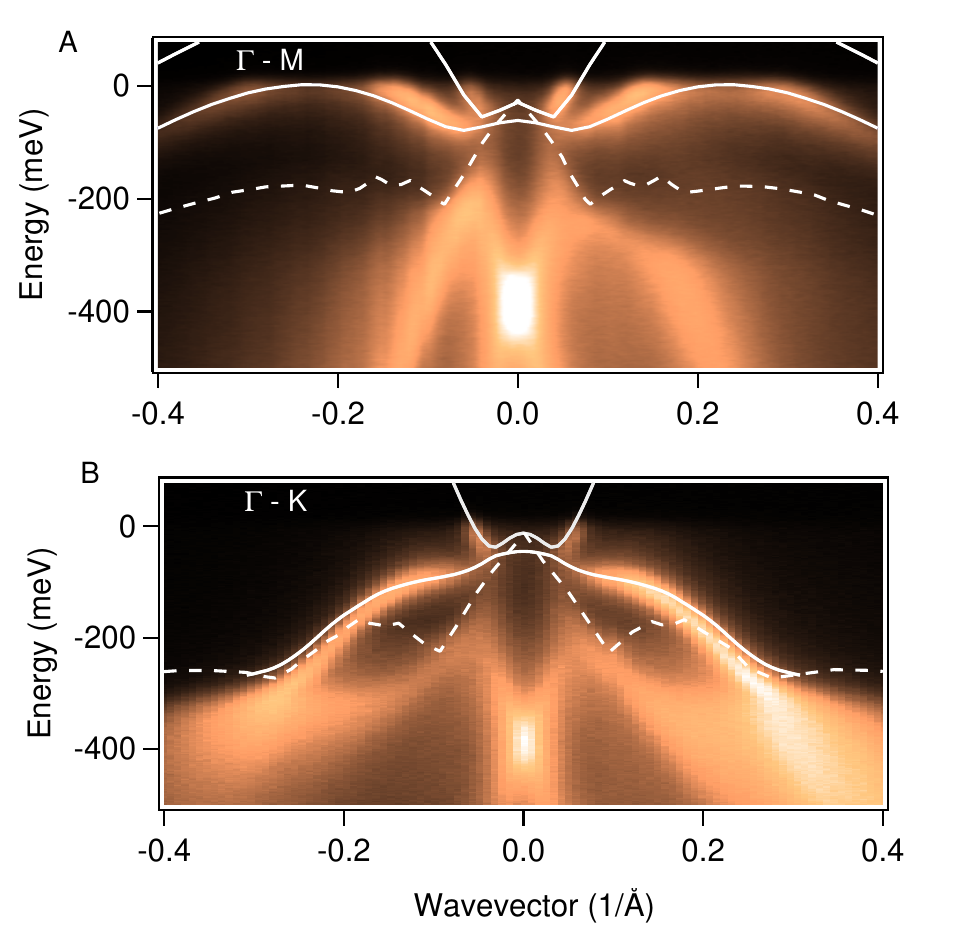}
\caption{A-B): Photoelectron intensity map acquired along the $\Gamma$-M direction (Panel A) and $\Gamma$-K direction (Panel B) is compared to the calculated band structure. The data have been acquired with circularly polarized photons centered at 18 eV. The solid lines stand for the two surface states whereas the dashed line marks the upper boundary of the projected bulk states that are below $E_F$.} \label{Fig2}
\end{center}
\end{figure}

In agreement with these results Fig. \ref{Fig2} shows that the SO interaction removes the Kramers degeneracy of the surface states
other than the $\Gamma$ point.
We recall that these wavefunctions hybridize with bulk states in the near proximity of the zone center while they are truly evanescent for larger wavevectors \cite{Koroteev1}. The data show that both spin-polarized bands cross the Fermi level along the $\Gamma$-M direction (panel A) whereas a single band crosses $E_F$ along $\Gamma$-K (panel B). In both cases, the higher lying band is above the Fermi level for parallel wavevector $0.07<k_{||}<0.5$ \AA$^{-1}$. Since these states are not accessible by standard ARPES experiments, the value of the SO splitting could be inferred only by the DFT calculations.

\begin{figure} \begin{center}
\includegraphics[width=1\columnwidth]{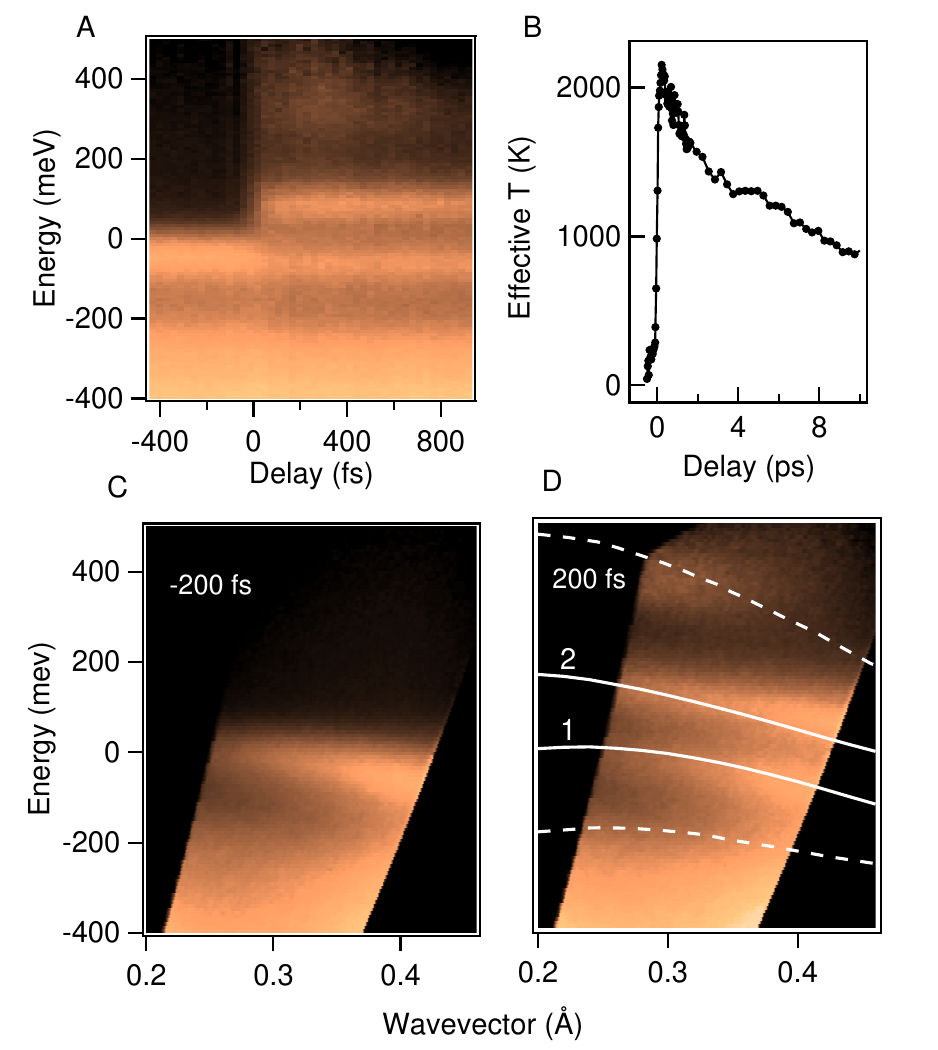}
\caption{A): Photoelectron intensity map as a function of pump-probe delay for $k_{||}=0.35$ \AA$^{-1}$ along the $\Gamma$-M direction. B): Temporal evolution of the effective electronic temperature. C-D): Photoelectron intensity map acquired along the $\Gamma$-M direction (red line in Fig. \ref{Fig1}) just before (C) and 200 fs after (D) the arrival of the pump beam. Solid lines stand for the calculated dispersion of the surface states while no projected bulk bands exist in between the two dashed lines. The data of this figure have been generated by ultrafast pulses of linearly polarized photons centred at 6.3 eV. The electric field polarization was nearly parallel to the surface plane and formed an angle $\alpha=20^\circ$ with respect to the Gamma-M direction.} \label{Fig3}
\end{center}
\end{figure}

\begin{figure} \begin{center}
\includegraphics[width=1\columnwidth]{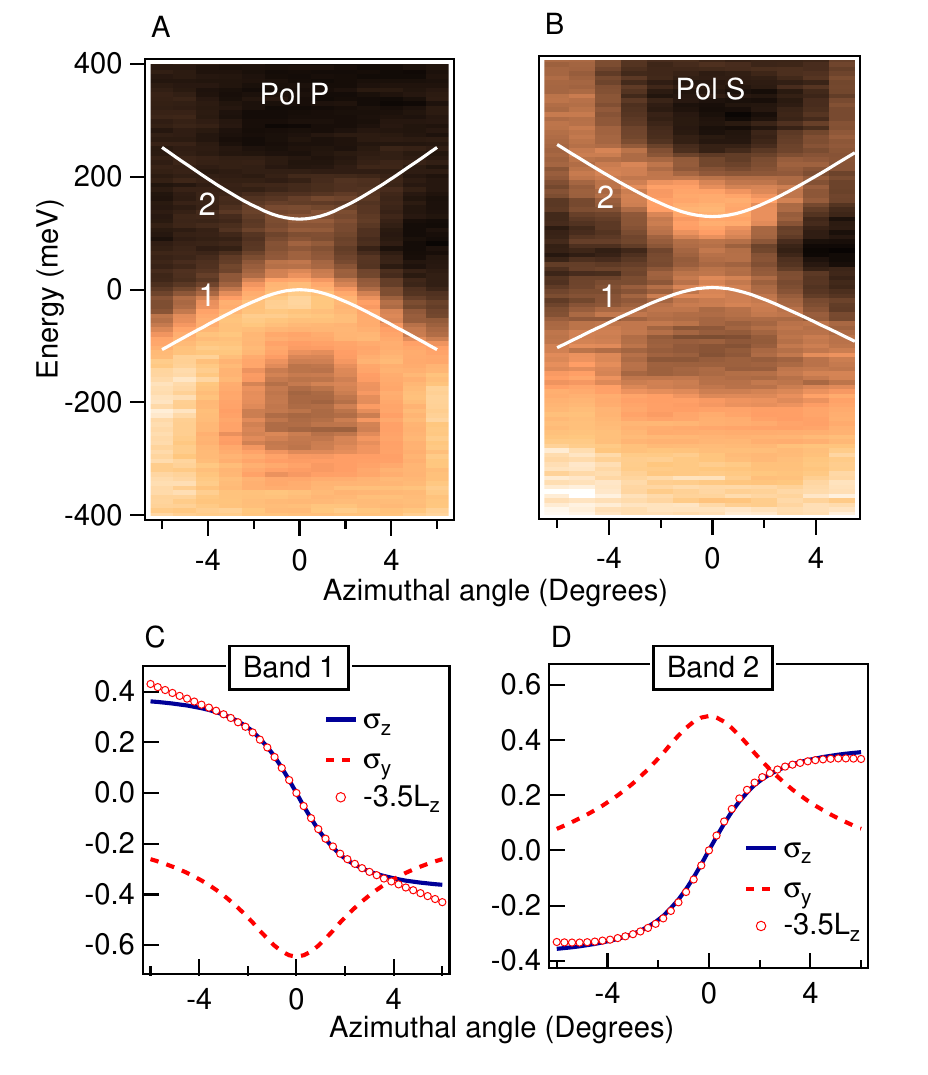}
\caption{A-B): Photoelectron intensity map as a function of azimuthal angle for $k_{||}=0.35$ \AA$^{-1}$ (blue line in Fig. \ref{Fig1}) and pump-probe delay of 200 fs. The probing photons are polarized either parallel (Panel A) or orthogonal (Panel B) to the $\Gamma$-M direction. The white lines stand for the calculated dispersion of the surface states. C-D): In-plane component of the spin ($\sigma_y$, red dashed line), out-of-plane component of the spin ($\sigma_z$, solid blue line) and of the orbital momentum (L$_z$, open circles) for the lower band (Panel C) and upper band (Panel D) as a function of azimuthal angle. The orbital momentum L$_z$ has been multiplied by -3.5 for better comparison with the out-of-plane spin.} \label{Fig4}
\end{center}
\end{figure}

In order to overcome this limitation, we measure the SO splitting by means of time resolved Photoelectron Spectroscopy. The bismuth surface is excited by an ultrafast pulse centered at 1.6 eV and is subsequently probed by a delayed 6.3 eV pulse. Figure \ref{Fig3}A shows the photoelectron intensity map acquired as a function of pump-probe delay for $k_{||}=0.35$ \AA$^{-1}$ along the $\Gamma$-M direction. The transient population of electronic states decays via emission of phonons on a timescale of 6 picoseconds. Since the electron gas thermalizes within 100 fs, the occupation factor is approximately described by a Fermi-Dirac distribution. Figure \ref{Fig3}B displays the temporal evolution of the electronic temperature after photoexcitation. More details on the procedure to extract this parameter can be found in the supplementary material of Ref. \cite{Papalazarou}.
Notice in Fig. \ref{Fig3}A that the SO-splitting does not depends appreciably on the pump-probe delay. We conclude that the potential barrier at the surface of bismuth is barely affected by the elevated electronic temperature. This finding is somehow unfortunate, as it precludes the possibility to modulate the SO coupling by light fields. Nonetheless, a spin dynamics may still be visible on the doped interface of Bi$_2$Se$_3$ \cite{King}. In this case, the electronic structure of quantum well states could change upon the photoinduced reduction of band bending.

Figure \ref{Fig3}C and \ref{Fig3}D shows the photoelectron intensity maps aquired along the $\Gamma$-M direction for a pump-probe delay  $\tau=-200$ fs and +200 fs, respectively. The observed SO splitting is almost constant in the measured $k_{||}$ interval. Our experimental estimate $\Delta_{\Gamma M}^{exp}=140\pm10$ meV is consistent with $\Delta_{\Gamma M}^{th}=130$ meV extracted from DFT calculations (white solid lines in Fig. \ref{Fig3}D).
The SO spitting is strongly reduced with respect to the atomic value but it is large for a device that would operate at room temperature. Notice in Figure \ref{Fig3}A and \ref{Fig3}D that the pump beam also populates a bulk derived band above the surface state. This structure at 370 meV nearly matches the lower boundary of the projected bulk bands that are above $E_F$ (upper dashed line).

In the following we focus on the giant anisotropy of the spin-orbit splitting.
We show in Fig. \ref{Fig4} the photoelectron intensity maps acquired at $k_{||}=0.35$ \AA$^{-1}$ and $\tau=200$ fs as a function of azimuthal angle $\varphi$. The probe pulse polarization was either parallel (panel A) or perpendicular (panel B) to the $\Gamma$-M direction. Notice that the lower and higher lying bands are more intense for parallel and perpendicular polarization, respectively. We conclude that the surface states have opposite parity with respect to the $\Gamma$-M plane. According to the selection rules of the dipole matrix elements the band below $E_F$ is even while the upper one is odd. The splitting between the two bands is minimal along the $\Gamma$-M direction but strongly depends on the azimuthal angle. When $\varphi$ reaches $\pm6^\circ$ the SO splitting attains 500 meV, thus increasing by 250\% with respect to $\Delta_{\Gamma M}^{exp}$. The DFT calculations display a similar trend but slightly underestimate the splitting value. Such large anisotropy is due to: a) the very large SO coupling of the atomic bismuth and b) the existence of surface states with large electronic wavevectors \cite{Claessen}. As a term of comparison, the second condition is not verified in Bi$_2$Te$_3$ and Bi$_2$Se$_3$, where the SO anisotropy induces only a minor warping of the Dirac cone \cite{Fu,Souma,Kuroda}.

The question arises as which spin texture originates from to the measured band structure.
On the $\Gamma$-M line, the mirror symmetry forbids the out-of-plane spin components and the spin polarization of the surface states fulfils those expected from the Rashba term $\alpha_R(k_y\sigma_x-k_x\sigma_y)$.
In contrast, away from the $\Gamma$-$M$ line, the anisotropy of splitting denotes a departure from a pure in-plane ordering. The leading correction is due to the cubic term $k_x(k_x^2-3k_y^2)\sigma_z$ and results from the in-plane deformation of the surface wavefunctions.

In order to check the spin ordering predicted by the measured electronic states, we projected the Kohn-Sham wavefunctions on the atomic orbitals of the bismuth atoms with defined $\sigma_z$ value ($z$ being the surface normal).
A realistic semi-infinite crystal has been mimicked by summating on the half side of the slab \cite{Hirahara}.
The expectation value of $\sigma_y$ is obtained as $\sigma_z$ but projecting on the spin polarization lying in-plane and perpendicular to $\Gamma$-$M$.
Figure \ref{Fig4}C and D show the spin polarization for the lower and upper surface state, respectively.
These simulations indicate that the out-of-plane polarization dominates the in-plane one as soon as $\phi>2{^\circ}$. In agreement with our finding, spin-resolved ARPES measurements reported a large out-of-plane buckling of the spins \cite{Takayama}.

Next, we discuss the entanglement taking place between the spin and orbital degrees of freedom near the atomic cores. We obtain the perpendicular component of the orbital polarization
by projecting of the Kohn-Sham wavefunctions on $p$ orbitals with angular momentum $l_z = \pm 1$. We verified that contributions from angular momentum larger than 1 are negligible.
As shown by Fig. \ref{Fig4}C and D, the perpendicular polarization of the spin and angular momentum are nearly proportional to each other. The scaling factor of -3.5 indicates that spins and orbital momenta are counter-aligned. This instance resembles the entanglement of $J=1/2$ orbitals in the strong coupling limit, albeit with electronic states carrying an orbital polarization quenched with respect to the atomic value. Indeed, neither the spin nor the orbital momentum is a good quantum number of the system. Despite it, the wavefunctions display an average orbital momentum $L_z$ that directly correlates to $\sigma_z$ \cite{Park}.

In conclusion, we investigated the bismuth (111) by time resolved photoelectron spectroscopy. The transient population of the electronic states above the Fermi level allows us to map the SO splitting in reciprocal space. We observed a giant anisotropy of the splitting value around the high symmetry planes. This finding is not compatible with a model based solely on the Rashba coupling and chiral spin texture. On the other hand the experimental dispersion of the electronic states is in good agreement with the results of first principle calculations. It follows that a large out-of-plane buckling of the spin and orbital polarization is essential to obtain the electronic structure of C$_{3v}$ surfaces with strong SO coupling. Such buckling is only a minor perturbation in proximity of the zone center but dominates the Rashba term at large electronic wavevectors.

We acknowledge that the FemtoARPES project was financially supported by the RTRA Triangle
de la Physique, the ANR program Chaires d'Excellence (Nr. ANR-08-CEXCEC8-011-01).


\begin{thebibliography}{99}
\bibitem{Note}
Discrepancies between DFT calculations and surface states dispersion has been noticed in proximity of the $M$ point. Nonetheless, the agreement between experiment and simulations is very good in the region of reciprocal space investigated by this work.
\end{thebibliography}
\end{document}